# Active control of coherent dynamics in hybrid plasmonic MoS$_2$ monolayers with dressed phonons

*Yuba Poudel[1], Gary N. Lim[2], Mojtaba Moazzezi,[1] Zachariah Hennighausen[3], Yuri Rostovtsev[1], Francis D'Souza[2], Swastik Kar[3], \*Arup Neogi[1]*

[1]Department of Physics, University of North Texas, Denton Texas, 76203, USA

[2]Department of Chemistry, University of North Texas, Denton Texas, 76203, USA

[3]Department of Physics, Northeastern University, Boston MA, 02115, USA

## ABSTRACT

The near-field interaction due to a strong electromagnetic field induced by resonant localized plasmons can result in a strong coupling of excitonic states or the formation of hybrid exciton-plasmon modes in quantum confined structures. The strength of this coupling can be increased by designing a system with its vibronic states resonant to the energy of the driving field induced by the localized plasmon excitation. Silver (Ag) nanoparticles (NPs) nucleated on molybdenum disulfide (MoS$_2$) is an ideal platform for such interaction. The influence of localized plasmons (LSP) on the formation and dissociation of excitons due to resonant and off-resonant optical excitation of carriers to excitonic states is studied using ultrafast optical spectroscopy. The presence of Ag-NPs generates a local field that enhances the magnitude of the Raman modes in MoS$_2$ under the resonant plasmon excitation. An ultrashort pulsed optical excitation at ~ 2.3 eV resonantly excites the LSP modes and the optical near-field resonantly drives the phonon modes, which leads to a coherent coupling of the A and B excitons in MoS$_2$ with the plasmon modes. The localized near-field optical driving source induces dressed vibronic states. The resonant excitation of the LSP modes modulates the optical absorption of the probe field. The optical excitation at ~ 3.0 eV, which is resonant to the C excitonic state but off-resonant to the LSP modes, increases the electrostatic screening in the presence of excess carriers from Ag-NPs. It results in a faster dissociation of optically generated C excitons into free carriers that eventually increases the population of A and B excitonic states. The coherent interaction in the hybrid nano-plasmonic system is described using a density matrix theory.

Two-dimensional layered transition metal dichalcogenides (TMDs) with quantum confinement and strong optical nonlinearities provide the ideal platform for designing photonic devices based on coherent light matter interaction.[1–6] Hybrid metal-semiconductor structures and plasmonic nanostructures are being increasingly used as a material platform to achieve strong coupling between photonic or plasmonic modes and electronic transitions.[7–13] These interactions are strongly enhanced in the presence of the localized optical field due to the plasmonic nanostructures and the quantum confinement of electrons and holes within lower dimensional semiconductors.[3,10,14] Carrier confinement in monolayer 2D materials such as graphene or TMDs exhibit high exciton binding energy and demonstrate a stable excitonic system at room temperature.[15,16] In these monolayer confined structures, the strong Coulomb effects also result in many-body effects such as charged exciton (trion) formation,[17] carrier induced broadening,[18] biexciton formation[19] and exciton annihilation[20]etc., which can significantly modify coherent interaction in strongly coupled systems.

Recent reports on exciton-plasmon coupling in TMDs system are mainly focused on achieving strong coupling in the steady-state regime.[1–4,6,21] However, the study of coherent dynamics in the strong coupling regime is significant as effects such as exciton-plasmon dephasing,[22] electronic and molecular charge transfer[23,24] and electromagnetically induced transparency[25] occur in the sub-picosecond domain and have potential applications in the development of high speed optoelectronic modulators, lasers and switches. Recently we have demonstrated electromagnetically induced transparency in the transient absorption domain due to plasmons coupled to excitons in graphene oxide quantum dots.[25] However, in a monolayer TMD semiconductors like $MoS_2$, the exciton dynamics are complex and can be strongly influenced by valence band splitting.[26] The electron and hole (e-h) pairs generated by the optical excitation at different energies result in the formation of A and B excitons. As these excitons are closely located in energy space, the two-excitonic states mutually drive correlated interactions within a few hundred femtoseconds (fs) to a picosecond (ps) time.[18] Additionally, the interband transition between the valence and conduction band can generate C excitons. An optical excitation at 3.06 eV causes an interband transition resonant to C exciton and creates holes deep in the valence band with a low probability of scattering to the K point. Consequently, it results in the generation of lower density of A and B excitons.

The dynamics of the three excitonic states in $MoS_2$ in the presence of a plasmonic mode makes it an attractive material system for the investigation of transient effects in the strong coupling regime. The coupling of plasmons to the excitons in the metal-$MoS_2$ system can be modified passively by selecting the proximity of the LSP energy with respect to the energies of A and B exciton. These interactions can be further modified by choosing an active process where the optical excitation energy can be tuned to control the excitonic transition or free-electron concentration due to interband transitions resonant or above the excitonic state. Due to the proximity of LSP energy of commonly used spherical gold (Au) NPs to the excitonic states of $MoS_2$, Au-$MoS_2$ hybrid structure has been commonly used to study the resonant interaction of localized plasmons with excitons in layered $MoS_2$ systems.[9] Ag thin film has a surface plasmon energy in free space is at about 3.05 eV, which is close to the interband transition energy of $MoS_2$ near Γ- point .[27] However, the surface plasmon energy of a metal thin film also depends on the dielectric constant of the substrate. As a result, the surface plasmon energy of Ag thin film on $MoS_2$ is reduced below 3.05 eV. In case of Ag-NPs, the corresponding localized plasmon energy strongly depends on the shape and size of the nanostructures. The nanoparticles considered in this work have been designed to have a localized plasmon energy centered around 2.33 eV, which is higher than the energy of A and B excitonic absorption states in $MoS_2$. The Raman modes of $MoS_2$ will be more strongly affected by an optical excitation resonant to the LSP energy at 2.33 eV compared to excitation at any other off-resonant frequency. The near-field interaction of the incident light scattered by the metal nanoparticles will drive the phonon modes more strongly at 2.33 eV compared to the Raman modes generated with an optical excitation at 3.06 eV in the absence of near-field interaction. The presence of A and B excitonic states close to the localized plasmon energy in the presence of strongly driven Raman field at 2.33 eV would further enhance the light-matter interaction in 2D materials.[25,28–31]

In this report, the ultrafast interaction of excitons with plasmons is studied in monolayer $MoS_2$ system using transient pump-probe optical spectroscopy in the presence of dressed phonon states generated by Ag-NPs. The dressed phonon modes are resonantly driven by the scattered optical near-field induced by the excitation of the Ag-NPs at 2.33 eV. The localized plasmon being broadband, the exciton dynamics without the resonant interaction are studied by using a higher energy optical excitation at 3.06 eV, which is resonant to C excitonic state. The exciton dynamics are significantly different in Ag-$MoS_2$ compared to bare $MoS_2$ monolayer due to the dressing of

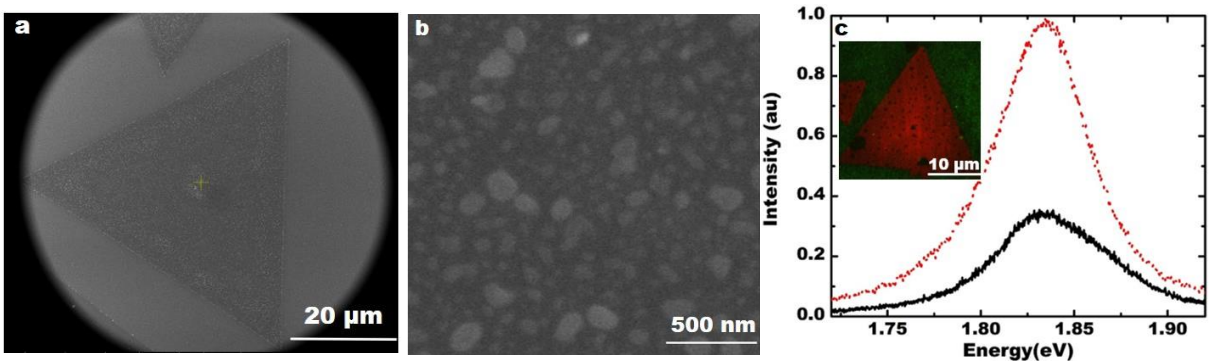

**Figure 1.** (a) FESEM image of Ag nanoislands on monolayer $MoS_2$. (b) Zoomed in view showing the Ag nanoislands distribution. (c) Enhanced photoluminescence (PL) from Ag-$MoS_2$ (dotted line) compared to $MoS_2$ (solid line) measured with 2.33 eV optical excitation. The inset shows the fluorescent image of $MoS_2$ with Ag-NPs

Raman modes. The plasmon mediated phonon process results in the modification of the inter-exciton coupling and generates coherently coupled exciton-plasmon hybrid state. A semiclassical density matrix model is used to illustrate the optical coherence during the light-matter interaction in the hybrid structure and the decay of the coherent excitonic interaction in $MoS_2$. The rate of the excitonic dephasing and the absorption saturation can be modulated by the active control of the pump light energy resonant to either the plasmon driven phonon energy at 2.33 eV or the C excitonic state at 3.06 eV. The coherent coupling of the excitons and plasmons due to the resonant excitation of the localized plasmons result in the generation of the A and B excitons as early as 210 fs which is not observed in the bare $MoS_2$ system.

**Structural Characteristics**

The structural characteristics of the hybrid Ag-$MoS_2$ system are shown in figure 1(a) and 1(b). The field emission scanning electron microscopic (FESEM) characteristics show large area monolayer $MoS_2$ on a quartz substrate with lateral dimensions extending over 50 microns. The distribution of Ag NPs on the $MoS_2$ layer is relatively non-uniform in terms of shape and size of the particles formed. A statistical analysis reveals that the average diameter of the Ag- NPs is in the range of 40 nm - 50 nm, although some bigger sizes were also observed. The presence of the Ag-NPs enhances the PL emission from the $MoS_2$ monolayer, as shown in figure 1(c), which is consistent with reported results.[3,9] The PL enhancement demonstrates that the presence of the Ag-NPs results in an enhanced spontaneous emission due to increased radiative recombination process. The inset in figure 1(c) is the fluorescent image of $MoS_2$ monolayer with Ag-NPs. The emission occurs in longer time scales and the origin of emission enhancement due to Ag-NPs will be discussed in another report.

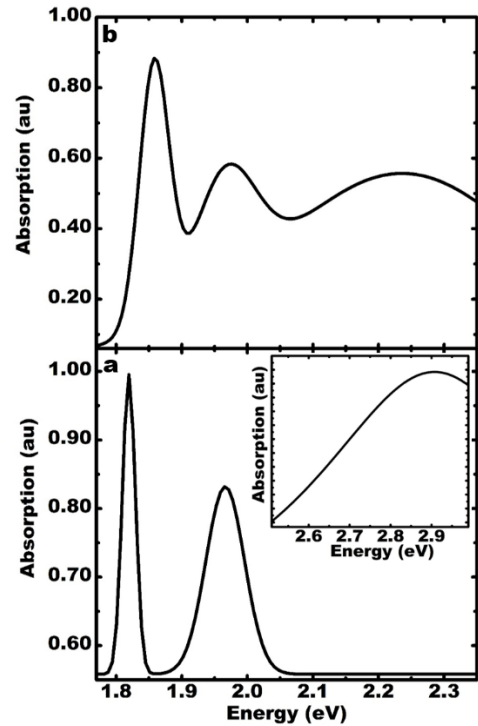

The room temperature excitonic absorption spectra were measured using a spectrophotometer with a white light source and a photomultiplier detector. The absorption spectrum of the bare $MoS_2$ layer shows two excitonic bands centered at 1.82 eV and 1.97 eV and represent the A and B excitonic states respectively, as shown in figure 2(a). In addition, a broad band centered at 2.90 eV, called C exciton, exists in the absorption spectrum, as shown in the inset. The absorption spectrum for the Ag-$MoS_2$ system is shown in figure 2(b).[32,33] The Ag-$MoS_2$ monolayer system depicts a distinct plasmon band centered at 2.25 eV for Ag-NPs. The resonances at 1.85 eV and 1.97 eV represent A and B excitons respectively. The larger dimension of the Ag-NPs is expected to shift the plasmon energy to the lower energy regime. Thus an optical excitation energy at 2.33 eV is expected to resonantly excite the localized plasmon modes in Ag-$MoS_2$, whereas a higher energy excitation at 3.06 eV should not result in any plasmon mode generation, but causes an interband transition in $MoS_2$.

**Figure 2.** (a) Absorption spectrum of $MoS_2$ monolayer showing the A and B excitonic absorption states in $MoS_2$ at 1.82 eV and 1.97 eV respectively. The inset shows a broad C exciton state centered at 2.90 eV. (b) Absorption states in the hybrid Ag-$MoS_2$. There is a localized plasmon band centered at 2.25 eV in the hybrid structure in addition to excitonic states.

The optical excitation resonant to plasmon mode in the hybrid Ag-$MoS_2$ system can induce the coupling of

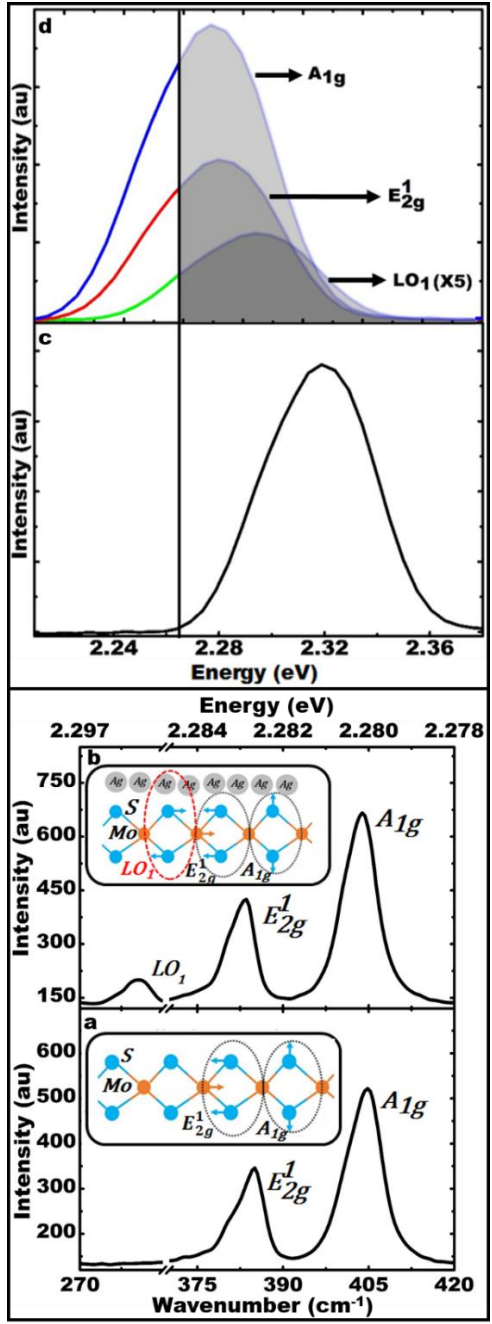

**Figure 3.** Raman spectrum of (a) MoS$_2$ and (b) Ag-MoS$_2$. The insets represents the corresponding vibrational modes. There are two active Raman modes - $E_{2g}^1$ and $A_{1g}$ in MoS$_2$. An additional Raman mode - $LO_1$ is observed in the hybrid structure. (c) Femtosecond pulse profile of the excitation source at 2.33 eV. (d) The estimated Raman bands in Ag-MoS$_2$ based on the Raman modes measured with 2.33 eV laser line. The shaded regions represent the relative overlap of the Raman bands with the fs laser pulse.

plasmon mode with the phonon modes in MoS$_2$. Thus the vibronic modes play important role in the exciton-plasmon interaction. The scattering spectrum from MoS$_2$ measured using micro-Raman spectroscopy with an optical excitation of 2.33 eV from a laser source is shown in figure 3(a). The inset shows the lattice vibrational modes in MoS$_2$. The observed position of $A_{1g}$ and $E_{2g}^1$ Raman modes confirms the MoS$_2$ layer to be a monolayer. A new Raman mode is detected at 280 cm$^{-1}$ in the hybrid structure that originates from $LO_1$ branch, as shown in figure 3(b), and the corresponding vibrational mode is shown in the inset.[28] This new Raman mode is attributed to symmetry breaking due to the near-field interaction with Ag plasmons. The active Raman modes are also shifted to lower wavenumber and are broadened in the hybrid structure due to the dressing of phonons in the presence of resonant excitation of the plasmons by a laser at 2.33 eV.[28] The fs laser pulse has a spectral full width at half maxima (FWHM) of about 50 meV at 2.33 eV, as shown in figure 3(c). Thus an optical excitation centered at 2.33 eV will drive all the three vibronic modes, in addition to be in resonance with localized plasmons, as shown in figure 3(d). On the other hand, the excitation at 3.06 eV does not drive the plasmons. Therefore the Raman modes are not driven by plasmon modes but still interact with the C excitons due to interband transitions. In the absence of any near-field radiation induced by the metal nanoparticles, the interaction occurs without an optical dressing of the incident photons by the phonon modes. At higher energies, there is a relatively weak coupling of phonons with incident light

**Theoretical Model**

The exciton dynamics in MoS$_2$ in the presence of local field induced by the plasmonic NPs is modeled.[34,35] This model takes into account self-consistently the interaction of excitons with LSP excited in the Ag-NPs to obtain a qualitative agreement with the observed pump-probe spectra and the spectral behavior at different delay times after excitation with the pump pulse. The incident laser field interacts with and induces dipole moments in Ag-NPs as well as in excitons in $MoS_2$. The exciton dynamics in $MoS_2$ in the presence of near-field interaction has been compared with the dynamics obtained in the absence of near-field interaction simply by exciting bare $MoS_2$ layer. The exciton energy levels are shown in figure 4. To understand the behavior of the system, we use a simplified level scheme (shown in the inset of figure 4(a)). The level $|a\rangle$ is excited state and the levels $|c_A\rangle$ and $|c_B\rangle$ represent ground states, such that the transitions $|a\rangle \rightarrow |c_A\rangle$ and $|a\rangle \rightarrow |c_B\rangle$ correspond to A and B excitonic transitions respectively. The levels $|b_A\rangle$ and $|b_B\rangle$ correspond to the vibrational levels. The vibrational frequency is larger than the corresponding energy due to thermal excitation at room temperature (RT); $\hbar\Omega_v > kT_0/\hbar$, so we consider that, at RT, the vibrational states are not excited. The pump laser pulses couple all ground levels with respective excited levels inducing the Raman coherence between the ground states and the vibrational states. The relaxation time for Raman induced coherence as estimated from the Raman modes is about 7 ps - 10 ps. The duration of the pump laser pulse is 100 fs, which is much shorter than the relaxation time of the LSP. Thus the LSP field excites the Raman coherence. The Raman induced coherence leads to the formation of so-called dark state, which have been analyzed using a density matrix theory. Here we consider simplified model with only three-levels. This model provides a clear picture of the formation of so-called dark states and the subsequent strong modification of the absorption spectra of the probe pulse.

The interaction Hamiltonian describing the interaction of excitons with external laser fields in the rotating wave approximation can be written as,

$$H = \hbar\,[\Omega_{1A}^* e^{i\omega_{ab_A}t}|b_A\rangle\langle a| + \Omega_{2A}^* e^{i\omega_{ac_A}t}|c_A\rangle\langle a| + adj.] + \hbar\,[\Omega_{1B}^* e^{i\omega_{ab_B}t}|b_B\rangle\langle a| + \Omega_{2B}^* e^{i\omega_{ac_B}t}|c_B\rangle\langle a| + adj.] \quad (1)$$

where $|b_A\rangle\langle a|, |c_A\rangle\langle a|, |b_B\rangle\langle a|$ and $|c_B\rangle\langle a|$ are the projection operators, $\omega_{ab_A}, \omega_{ac_A}, \omega_{ab_B}$ and $\omega_{ac_B}$ are the exciton transitions that are changing in time due to exciton relaxation and these changes are taking into account phenomenologically,

$$\Omega_1 = \wp_{ab_\alpha}(\varepsilon_p e^{i\omega_p t} + \varepsilon_d e^{i\omega_d t})/\hbar \quad (2)$$

$$\Omega_2 = \wp_{ac_\alpha}(\varepsilon_p e^{i\omega_p t} + \varepsilon_d e^{i\omega_d t})/\hbar \quad (3)$$

are the Rabi frequencies, and $\wp_{ab_\alpha}$ and $\wp_{ac_\alpha}$ are the dipole moments of the transitions in A and B excitonic states correspondingly ($\alpha = A, B$), $\varepsilon_p$ and $\varepsilon_d$ are the probe and pump fields. A continuum white-light source has been used as probe to monitor the spectral evolution with respect to time, and the analysis of the transient spectrum has been considered for a range of excitonic transitions. The probe laser pulse can be delayed with respect to the pump laser pulse. A strong field of frequency $\omega_d$ is the pump field and a weak field of frequency $\omega_p$ has been considered as the probe laser field. The pump field drives the localized plasmons. Indeed the polarization of the nanoparticle related to the local electric field $E_{loc}$ is given by $P = \frac{(\epsilon-1)}{4\pi}E_{l0C}$, which has the time dependence that is described in the $\tau$-approximation by the equation,

$$\frac{\partial P}{\partial t} = -\gamma \left( P - \frac{(\epsilon - 1)}{4\pi} E_{loc} \right) \quad (4),$$

where γ is the electronic relaxation rate in nanoparticle, and $E_{loc} = E_0 + \frac{4\pi}{3} P$. Thus, we can see that the relaxation rate for the polarization is given by, $\gamma_{Fld} = \gamma \left( \frac{\epsilon+2}{3} \right)$ and at the surface plasmon resonance $\epsilon + 2 = 0$, which is longer than relaxation for electrons. The field created by localized plasmons drives the excitonic lines via the vibronic state with detuning $\Delta_{1\alpha} = (\omega_{ab_\alpha} - \omega_d)$, $\Delta_{2\alpha} = (\omega_{ac_\alpha} - \omega_d)$ and the probe fields are in resonance with detuning $\Delta_{p\alpha} = (\omega_{ab_\alpha} - \omega_p)$. The density matrix equations are given by,

$$\frac{\partial \rho}{\partial t} = \frac{i}{\hbar} [\rho, H] - \hat{\Gamma}[\rho] \quad (5),$$

where $\hat{\Gamma}[\rho]$ is the matrix of relaxation rates for all components of the density matrix $\rho$ that was taken into account phenomenologically. The incident laser field interacts with Ag-NPs as well as with the excitons in the nearest vicinity. The dipole moment is induced in Ag-NPs by the action of both fields; the incident laser field and the field that is created by surrounding excitons. The dipoles induced in excitons are created by the action of incident laser field and the field created by Ag-NPs.

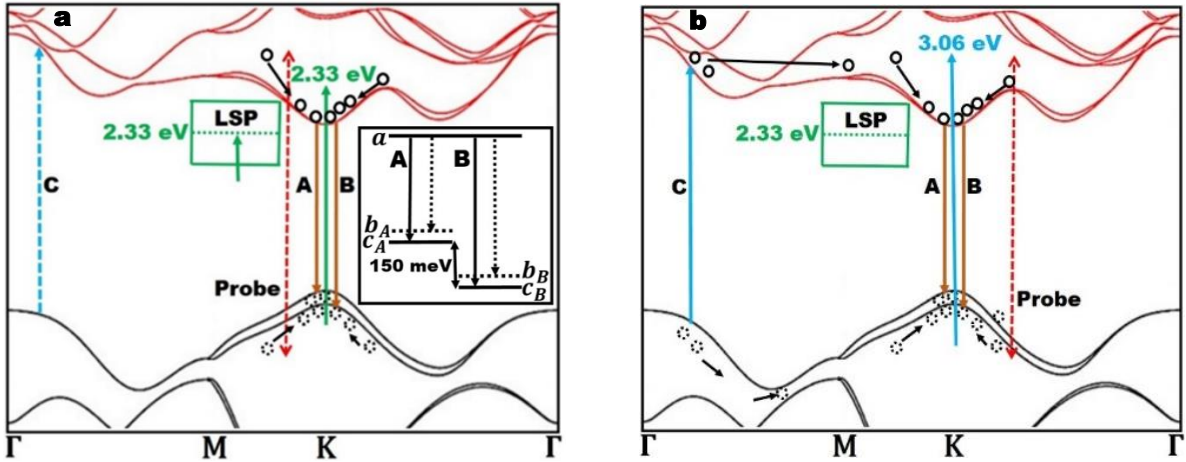

**Figure 4.** Energy band diagram of monolayer MoS$_2$ showing the optical excitation and the relaxation of carriers to excitonic states. The solid vertical line with upward arrow represents pump and the vertical dotted line with double sided arrows represent the probe. The LSP band is also shown. (a) With 2.33 eV excitation, the carriers are excited at K-point in MoS$_2$ with the pump pulse initially at higher energy states, which then relax to A and B excitonic states. The LSP mode is excited in Ag-MoS$_2$, which in turn coherently couples with the excitonic states. The inset represents the schematics of the density matrix model, The level $a$ is the excited state and the level $c_A$ and $c_B$ represent ground states such that the transitions $a \to c_A$ and $a \to c_B$ correspond to A and B excitons respectively. The levels $b_A$ and $b_B$ correspond to the vibrational levels and the Raman transitions are represented by dotted arrows. (b) The optical excitation at 3.06 eV can not generate the plasmon modes; however can excite the carriers at diffeent regions, which subsequently relax to A and B excitonic states through different intraband relaxations.

The set of density matrix equations are given by,

$$\dot{\rho}_{ab_\alpha} = -(\Gamma_{ab_\alpha} + i\Delta_{1\alpha}) \rho_{ab_\alpha} + in_{ab_\alpha}\Omega_{1\alpha} - i\Omega_{2\alpha}\rho_{c_\alpha b_\alpha} \quad (6)$$

$$\dot{\rho}_{c_\alpha a} = -(\Gamma_{c_\alpha a} - i\Delta_{2\alpha}) \rho_{c_\alpha a} + in_{c_\alpha a}\Omega_{2\alpha} + i\Omega_{1\alpha}\rho_{c_\alpha b_\alpha} \quad (7)$$

$$\dot{\rho}_{c_\alpha b_\alpha} = -(\Gamma_{c_\alpha b_\alpha} + i(\Delta_{1\alpha} - \Delta_{2\alpha}))\rho_{c_\alpha b_\alpha} + i(\rho_{c_\alpha a}\Omega_{2\alpha} - i\Omega_{1\alpha}\rho_{ab_\alpha}) \quad (8)$$

$$\dot{\rho}_{b_\alpha b_\alpha} = i(\rho_{b_\alpha a}\Omega_{1\alpha} - \Omega_{1\alpha}^*\rho_{ab_\alpha}) \quad (9)$$

$$\dot{\rho}_{c_\alpha c_\alpha} = i(\rho_{c_\alpha a}\Omega_{2\alpha}^* - \Omega_{2\alpha}\rho_{c_\alpha a}) \quad (10)$$

where $n_{ab_\alpha} = \rho_{aa} - \rho_{b_\alpha b_\alpha}$ and $n_{c_\alpha a} = \rho_{c_\alpha c_\alpha} - \rho_{aa}$ ; $(\alpha = A, B)$

Additional broadening of excitonic transition includes the interaction with metallic nanoparticles. The Raman induced coherence is given by,

$$\dot{\rho}_{c_\alpha b_\alpha} = i(\rho_{c_\alpha a}\Omega_1 - \Omega_2\rho_{ab_\alpha}) - (\Gamma_{c_\alpha b_\alpha} + i\Delta_{1\alpha} - i\Delta_{2\alpha} + \frac{|\Omega_{1\alpha}|^2}{(\Gamma_{c_\alpha a} - i\Delta_{2\alpha})} + \frac{|\Omega_{2\alpha}|^2}{(\Gamma_{ab_\alpha} + i\Delta_{1\alpha})} + i\Gamma_{NP})\rho_{c_\alpha b_\alpha} \quad (11)$$

where $\Gamma_{NP} = \alpha_p \Gamma_{dd}^{ab_\alpha} - \alpha_d \Gamma_{dd}^{c_\alpha a}$

and $\Gamma_{dd}^{c_\alpha b_\alpha} = \frac{\beta \wp_{c_\alpha b_\alpha}^2}{\hbar}$, $\Gamma_{dd}^{ab_\alpha} = \frac{\beta \wp_{ab_\alpha}^2}{\hbar}$ and $\Gamma_{dd}^{c_\alpha a} = \frac{\beta \wp_{c_\alpha a}^2}{\hbar}$ are the effective coupling of excitons with Ag-NPs, $\beta$ is related to the average distance between exciton and the Ag-NP; such that the quantities $\beta \wp_{ab}^2$ and $\beta \wp_{ac}^2$ are the characteristic dipole-dipole interaction between excitons and Ag-NPs, $\alpha_d = \tilde{\alpha}(\omega_d)$, $\alpha_p = \tilde{\alpha}(\omega_p)$ and

$$\tilde{\alpha}(\omega) = -2\frac{\epsilon_1(\omega) - \epsilon_2}{\epsilon_1(\omega) + 2\epsilon_2} \quad (12)$$

where $\epsilon_2(\omega)$ is the dielectric function of the Ag-NP, ω is the frequency of the electric field, $\epsilon_2$ is the dielectric function of the medium. It is important to note here that there is pump field dependent term - $(\frac{|\Omega_{1\alpha}|^2}{(\Gamma_{c_\alpha a} - i\Delta_{2\alpha})} + \frac{|\Omega_{2\alpha}|^2}{(\Gamma_{ab_\alpha} + i\Delta_{1\alpha})})$ in equation (11). This term determines the timescale for the excitation of the coherence due to pump field. This term also leads to the modification of the excitonic relaxation rate due to Raman process induced coherence as described in the next section.

For the weak probe field, the set of density matrix equations can be solved adiabatically. We can see that each spectral component of the probe field induces corresponding Raman induced coherence.[36]

The spectrum of induced exciton polarization of the system is given by,

$$P_\omega = \sum_{\alpha=A,B} \wp_{ab_\alpha} \rho_{ab_\alpha,\omega} + \wp_{ac_\alpha}\rho_{ac_\alpha,\omega} \quad (13)$$

where

$$\rho_{ab,\omega} = -i\frac{\wp_{ab_\alpha}n_{ab_\alpha}(t) + \wp_{ac_\alpha}\rho_{c_\alpha b_\alpha}(t)}{\hbar(\Gamma_{ab_\alpha,\omega} + i\Delta_{p\alpha})} \varepsilon_p^\omega \quad (14)$$

$$\rho_{ca,\omega} = -i\frac{\wp_{ac_\alpha}n_{c_\alpha a}(t) - \wp_{ab_\alpha}\rho_{c_\alpha b_\alpha}(t)}{\hbar(\Gamma_{c_\alpha a,\omega} - i\Delta_{p\alpha} + i\Delta_{1\alpha} - i\Delta_{2\alpha})} \varepsilon_p^\omega \quad (15)$$

where $n_{ab_\alpha}(t) = \rho_{aa}(t) - \rho_{b_\alpha b_\alpha}(t), n_{c_\alpha a}(t) = \rho_{c_\alpha c_\alpha}(t) - \rho_{aa}(t), \rho_{aa}^{(t)}, \rho_{b_\alpha b_\alpha}^{(t)}$ and $\rho_{c_\alpha c_\alpha}^{(t)}$ are the population in the levels $|a\rangle, |b_\alpha\rangle$ and $|c_\alpha\rangle$ respectively that are obtained by solving the time dependent density matrix equations with the corresponding relaxation rates of the population of excitonic levels as well as the decay of coherence. The coupling of excitons with Ag-NPs induces the coherence $\rho_{c_\alpha b_\alpha}$ that strongly modifies the absorption spectra of the probe pulses.[36]

This model is used to simulate the exciton dynamics with and without the plasmonic interaction with the pump laser field at ~ 2.33 eV. In the absence of Ag-NPs, the excitons are driven by the pump field without the near-field interaction due to the surface charges from the localized plasmons. The calculations have been done using the parameters corresponding to the experimental conditions. The typical values used for the simulations are: $\Gamma_{ab_\alpha} \sim \Gamma_{ac_\alpha}$ = 152 ps$^{-1}$, $\gamma_A$ = 0.03 ps$^{-1}$, $\gamma_B$ = 0.05 ps$^{-1}$, localized plasmon relaxation rate $\gamma_{Fld}$ = 1.6 ps$^{-1}$ and $\Omega_{d_\alpha}$ = 320 ps$^{-1}$, the maximum Rabi frequency corresponding to plasmon near-field. All population are initially in the states $|c_B\rangle$ and $|c_A\rangle$, $\rho_{aa}(t=0) = 0$, $\rho_{c_A c_A}(t=0) = \rho_{c_B c_B}(t=0) = 0.5$, and no coherence is excited at t = 0, $\rho_{ab_\alpha}(t=0) = \rho_{b_\alpha b_\alpha}(t=0) = \rho_{c_\alpha b_\alpha}(t=0) = \rho_{c_\alpha a}(t=0) = 0$.

**Effect of optical excitation energy on the exciton dynamics of MoS$_2$**

The optical excitation at 2.33 eV initially excites the carriers to higher energy states. About 60 % of these free carriers generated by the pump pulse subsequently relax to respective excitonic states at the K- point through different intraband relaxation processes in sub-picosecond timescale.[37,38] The filling of excitonic states with the carriers gives rise to a decrease in absorption of the probe resulting in the formation of negative amplitude bands, called photobleaching (PB) bands, that appear in the transient absorption spectrum, as shown in figure 5. The spectral analysis shows that the PB bands consist of two Gaussian peaks fitted at the position of A and B excitons, as shown in supporting figure S2. Thus the amplitude of each PB band at the respective position of A and B excitons is a measure of the excitonic population in that state. The excitonic population is saturated in both states at a delay time of about 750 fs. A comparison of the exciton generation at 2.33 eV and 3.06 eV is shown in figure 5. The optical excitation at 3.06 eV initially generates C excitons near Γ- point as well as excites the carriers to higher energy states at K-point.[38] The dominant part of the imaginary dielectric function in monolayer MoS$_2$ at the position of C exciton results in a strong optical absorption that is approximately three times the absorption at the position of A and B excitons.[39–41] As a result, the excited carrier density is higher. At high carrier density, the Coulomb interaction between carriers is screened so that the generated electrons and holes act as free carriers.[42,43] The electrons rapidly scatter toward conduction band minimum at K- point.[44] However, there is a relatively lower probability for the holes to be scattered to the K- point in the valence band, so the holes still remain near Γ- point. Thus the density of A and B excitons formed is lower and the exciton dynamics are influenced by the free carrier–exciton interaction.[40,45] There is a stronger state filling with electrons in the conduction band that also contributes to the formation of the PB bands at the position of A and B excitons. There is higher PB at the position of B exciton due to state filling compared to that of A exciton. The PB bands saturate at a delay time of about 500 fs, which is much faster than with 2.33 eV excitation. This is a consequence of dissociation of C exciton and relaxation of electrons to conduction band minima.[18,44,46] The excited carriers in the longer-lived states absorb the probe, which gives rise to the formation of absorption band, called photoinduced absorption band, in the transient absorption spectrum. The photoinduced absorption bands evolve at different delay times and at different energy states. A broad photoinduced absorption band centered at 2.45 eV appears with the 3.06 eV excitation as early as a delay time of 210 fs. This photoinduced absorption band is attributed to the excitonic resonance.[47] The amplitude of this band gradually increases and saturates at a delay time of about 5.6 ps. However, with 2.33 eV excitation, photoinduced absorption bands are weaker as the photoexcited carriers have a higher probability to be scattered to the excitonic states. A weak photoinduced absorption band appears at 2.5 eV that saturates at a delay time of about 3 ps. The population of A and B excitonic states decreases faster over time with 2.33 eV excitation compared to the excitation at 3.06 eV. This can be attributed to the excitonic recombination, which eventually gives rise to the fast recovery of the absorption of the probe.

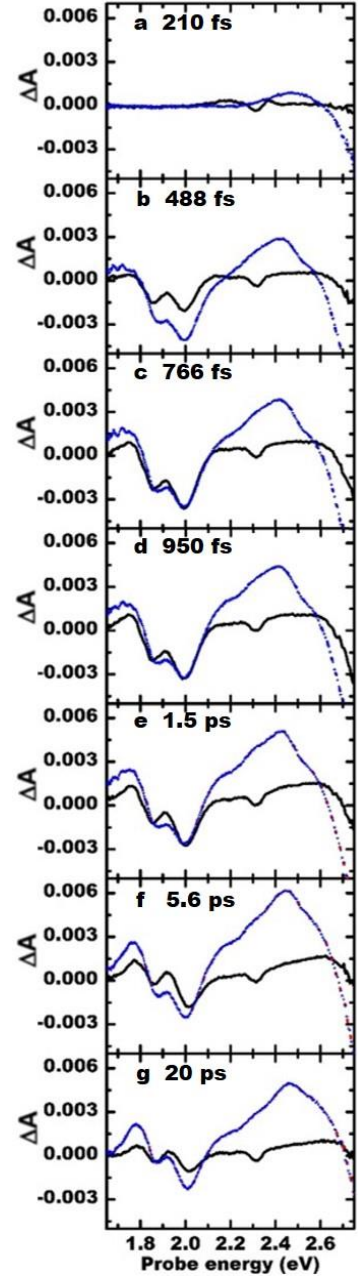

**Figure 5.** The Transient absorption spectrum of monolayer MoS$_2$ at different delay times with 2.33 eV (solid line) and 3.06 eV (dotted line with circle) optical pump energy respectively and probed with white light pulses. PB bands appear at the position of A and B excitonic states at a delay time of 488 fs. The amplitude, line width and position of PB bands vary with excitation energy. A strong and broad photoinduced absorption band centered at 2.5 eV appears with 3.06 eV excitation.

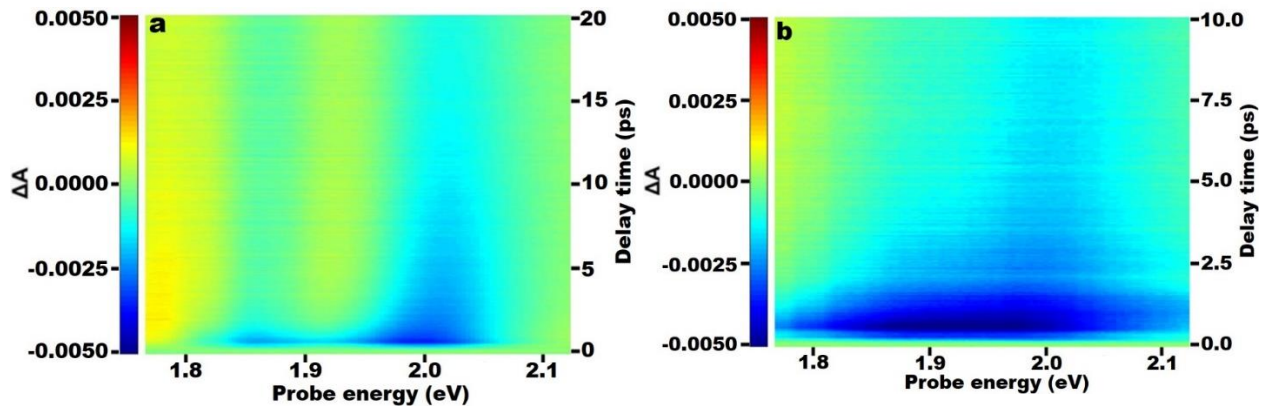

**Figure 6.** The transient absorption spectrum of (a) $MoS_2$ and (b) $Ag-MoS_2$ at different delay times with 2.33 eV pump energy and probed with white light. Strong PB bands appear at the position of A and B excitonic absorption states in $MoS_2$ at very short delay times. The absorption of the probe gets gradually recovered at longer delay times. In $Ag-MoS_2$, the excitonic absorption states are hybridized together with plasmon band forming a broadband around A and B excitonic states.

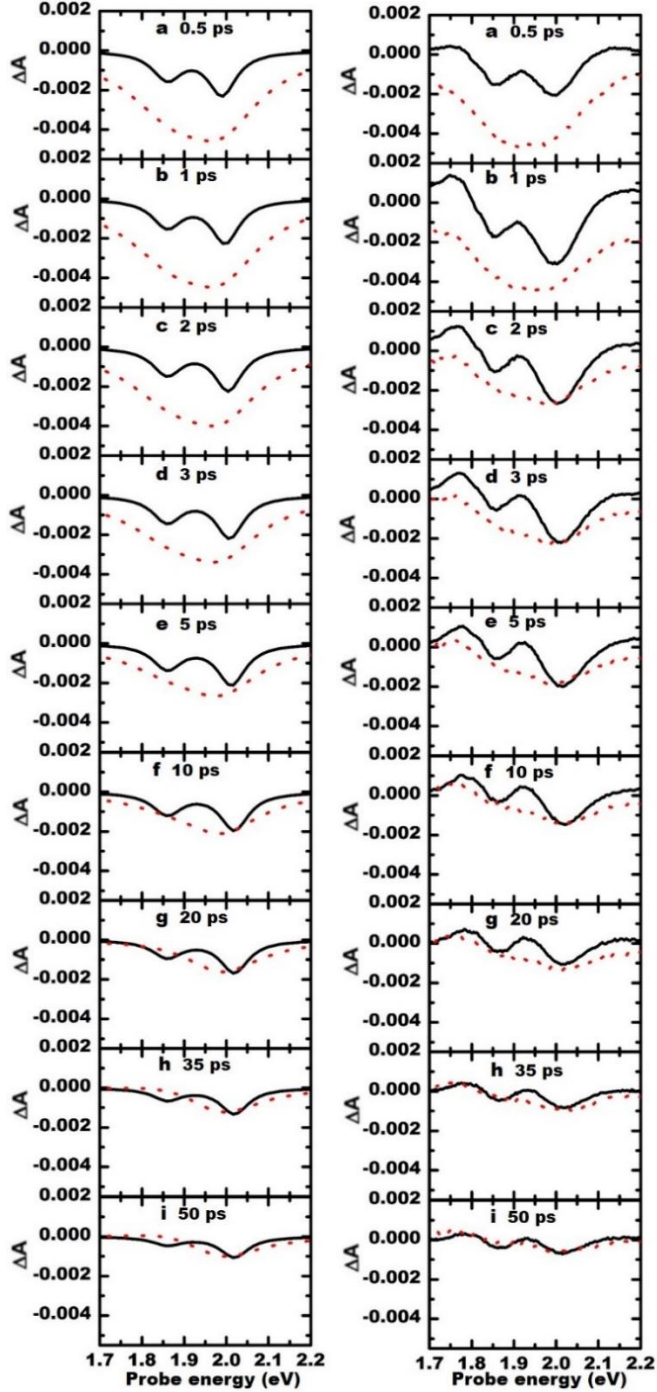

**Figure 7.** The spectral evolution of the transient absorption spectrum of monolayer MoS$_2$ (solid line) and Ag-MoS$_2$ (dotted line) with 2.33 eV excitation. The left and right panels respectively represent experimental spectra and calculated spectra using the density matrix model. The dynamics of the A and B excitonic bands at different delay times ranging from 0.5 ps to 50 ps, as indicated, in MoS$_2$ and the hybridization of the excitonic bands with the plasmon band in Ag-MoS$_2$ at short delay times as measured experimentally agree well with the corresponding dynamics of the calculated spectra.

## Exciton–plasmon coupling with resonant plasmon excitation

Figure 6 shows the evolution of the differential absorption spectrum at different delay times for an optical excitation with energy at 2.33 eV that is in resonance to the plasmon modes and simultaneously excites the Raman modes in the hybrid structure. The left and right panels in figure 7 respectively represent the spectral evolution of transient absorption spectra of MoS$_2$ and Ag-MoS$_2$ measured experimentally at different delay times and the corresponding transient absorption spectra calculated using the density matrix model. There is overall a good agreement between the experimental and the calculated spectra. The slight discrepancy in the dynamics of the calculated spectra is due to the fact that the many-body effects including the inter-exciton coupling between A and B excitons have not been included in this model. In the hybrid structure, the optical excitation results in the excitation of resonant plasmon modes in addition to the excitation of free carriers in MoS$_2$. There is an increased absorption in the hybrid structure induced due to enhanced electric field of resonantly excited plasmons. A summary of figures 6 and 7 shows the relative excitonic population change due to resonantly coupled plasmon modes to phonon modes with 2.33 eV excitation, as shown in figure 8. At an early delay time of about 210 fs, there is no change in the absorption of the probe pulse indicating that the carriers excited by the pump pulse have not relaxed yet to excitonic states in MoS$_2$. However, in the hybrid structure, the A and B excitonic states are already populated, as shown in the inset of figure 8 (b). The population of excitons is higher in the hybrid structure and saturates at a delay time of about 500 fs in the presence of driving filed generated due to plasmons. As described in the theoretical section, the near-field interaction in the presence of Ag-NPs drives the vibronic or Raman modes. Furthermore, the driving field creates coherence between ground levels. It is the ground state coherence ($\rho_{bc}$) that is responsible for such remarkable modification of the probe absorption. Without Ag-NPs, we observe an exponential recovery of the probe absorption, as shown in figure 8(a). However, in the Ag-$MoS_2$ system, there is pump field dependent term - $(\frac{|\Omega_{1\alpha}|^2}{(\Gamma_{c_\alpha a} - i\Delta_{2\alpha})} + \frac{|\Omega_{2\alpha}|^2}{(\Gamma_{ab_\alpha} + i\Delta_{1\alpha})})$ in equation (11). This term determines the timescale for the excitation of the coherence due to pump field. This term also leads to the modification of the excitonic relaxation rate due to Raman induced coherence. As a result, there is a faster decrease in the absorption of probe due to forming the so-called excitonic dark state as well as faster recovery of the probe absorption in the presence of pump field due to the presence of Ag-NPs, as shown in figure 8(b). Once the plasmonic field is weakened, the probe absorption recovers with exciton relaxation rate, similar to the decay of probe absorption without Ag-NPs. This behavior of the probe absorption can't be simply due to the relaxation of excitonic population influenced due to Ag NPs. The driving field causes power broadening of the exciton transitions. The additional broadening of the pump-probe spectra is caused by the spatial distribution of the excitons near Ag-NPs. This broad differential absorption band is the first experimental observation of hybrid exciton-plasmon band or the 'plexciton band' in the hybrid Ag-MoS$_2$ system measured using ultrafast spectroscopy. The plexciton band shows the initial blue shift due to presence of the Raman driven field and splits into separate A and B excitonic bands. A spectral peak fitting analysis of MoS$_2$ and Ag-MoS$_2$, as shown in supporting figure S2 and S3, gives a correlation of the decoherence time of the relaxation of the plexciton mode. At a delay time of about 700 fs to a ps, the plexciton band dissociates into separate A and B excitonic bands, as shown in figure 9. The A exciton band red shifts, whereas the B exciton band blue shifts immediately after the dissociation. The small energy difference between A and B excitons immediately after the dissociation is attributed to the change in exciton binding energy. These excitons tend to approach the energy of the unscreened excitonic energy states observed in MoS$_2$ without Ag-NPs. The A exciton recovers the energy levels at a faster rate than the B exciton. The presence of the driving field and the interaction with the localized plasmon field increases the coupling between the A and B excitons in the presence of the Ag-NPs.

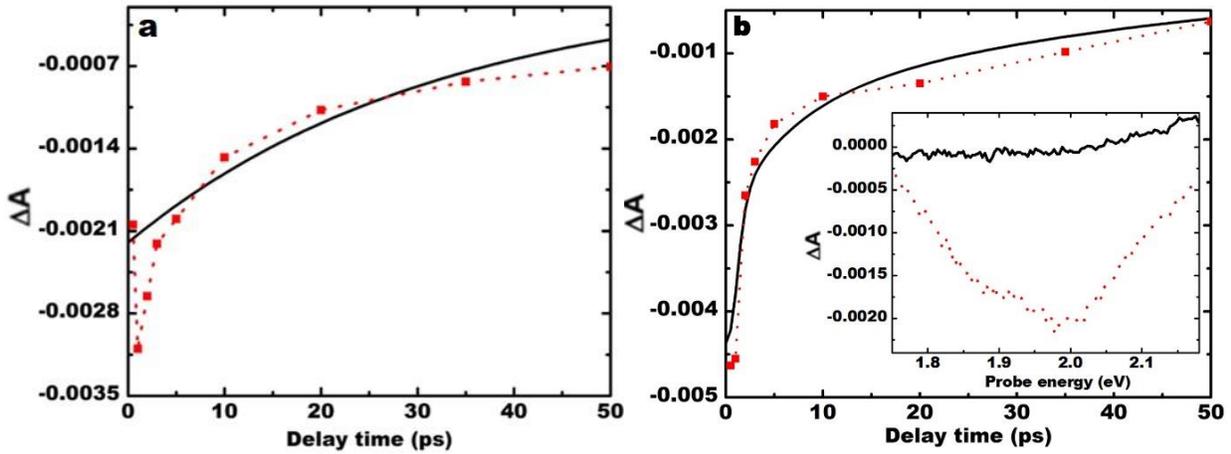

**Figure 8**. Comparison of theoretical (solid line) and experimental (dotted line with rectangles) decay characteristics of excitonic population in (a) $MoS_2$ and (b) $Ag-MoS_2$ showing the recovery of the absorption of probe. In $MoS_2$, the exciton formation rate exceeds the exciton dissociation rate up until a delay time of 750 fs. The absorption of probe gradually recovers at longer delay times. However, in $Ag-MoS_2$, the exciton population saturates and starts decaying at a delay time of 0.5 ps. The recovery of absorption of probe is faster in $Ag-MoS_2$ at short delay times and approaches to that in $MoS_2$ in few ps. The inset in figure b shows the formation of exciton bands in $Ag-MoS_2$ (dotted line) as early as 210 fs, which is not observed in $MoS_2$ (solid line).

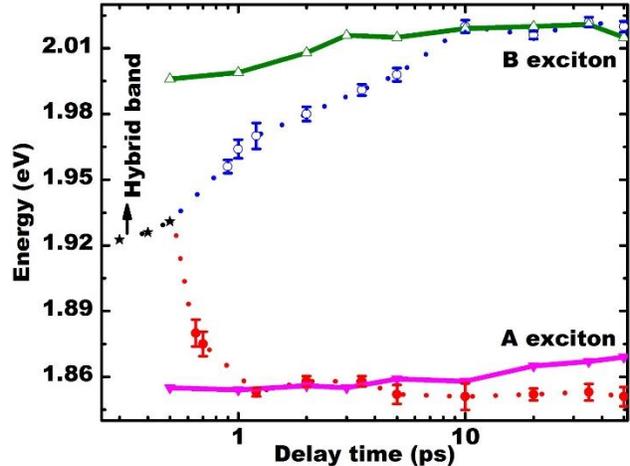

**Figure 9.** Dynamics of the hybrid exciton-plasmon band under the plasmonic field. The hybrid band (dotted line with solid star) initially blue shifts and splits into A exciton (dotted line with solid circles) and B exciton (dotted line with hollow circles) bands at a delay time of about 0.7 ps. The vertical lines with symbols represent the corresponding error in energy. The A and B excitonic bands in bare $MoS_2$ (solid lines with triangles) are shown for comparison.

**Modification in exciton dynamics under C exciton resonance**

Figure 10 shows the temporal and spectral evolution of the differential transient absorption spectrum with 3.06 eV in MoS$_2$ before and after the formation of the hybrid structure. The low exciton density in MoS$_2$ results in a weaker exciton-exciton interaction, which is indicated by a relatively smaller peak separation of 110 meV between the PB bands at A and B excitons. The PB bands in the hybrid structure saturate approximately at same delay time when probed at B exciton, but has a slightly longer saturation time at A exciton, than that in MoS$_2$. This can be attributed to the influence due to Ag-NPs. There is an increased electrostatic interaction between the optically generated carriers in the presence of Ag nanoislands that generates a higher density of A and B excitons, which results in an enhanced interaction between excitons and the separation increases to about 150 meV.[18] Initially, there is a higher excitonic population in Ag-MoS$_2$ due to enhanced relaxation rate of carriers to excitonic states, however, at a delay time of about 3 ps, the excitonic population drops to lower than that in $MoS_2$. This can be attributed to the modulated lifetime of the excitonic states in the presence of Ag nanoislands. There is stronger bandgap renormalization in Ag-$MoS_2$, resulting in a higher red shift of excitonic states compared to 2.33 eV excitation in the presence of excess free carriers.[47,48] A new PB band originates in Ag-$MoS_2$ at 2.56 eV at short delay times. This is attributed to the enhanced electrostatic interaction between optically generated carriers in the presence of Ag-NPs. The amplitude of this peak drops fast as the C exciton has a relatively short lifetime.[42] The strong photoinduced absorption band in $MoS_2$ at 2.45 eV is weakened due to enhanced relaxation of the excited carriers through additional relaxation pathways introduced due to Ag-NPs. The amplitude as well as the linewidth of photoinduced absorption band saturate at a delay time of about 5.6 ps as shown in supporting figure S1.

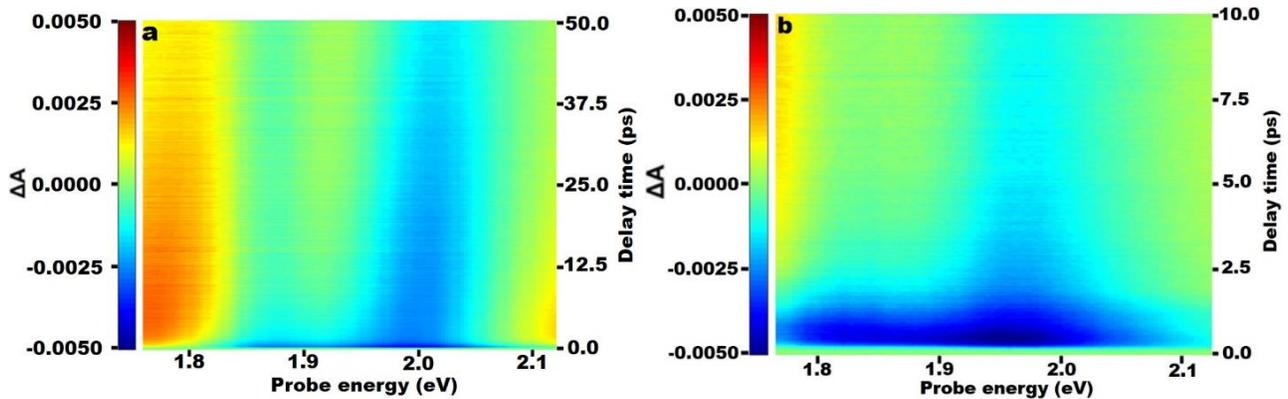

**Figure 10.** The transient absorption spectrum of (a) monolayer MoS$_2$ and (b) Ag-MoS$_2$ at different delay times with 3.06 eV pump energy and probed with white light. The PB bands at A and B excitonic absorption states are weaker in MoS$_2$ and are at a separation of about 110 meV. The PB bands got stronger and the spacing between the bands increases to 150 meV in Ag-MoS$_2$. The PB bands are red shifted in Ag-MoS$_2$ compared to MoS$_2$.

**CONCLUSIONS**

We described the influence of Ag-NPs on the exciton dynamics of $MoS_2$ under two different optical excitations that are resonant to either plasmon mode or the C exciton. The optical excitation resonant to plasmon mode drives the phonon modes due to near-field interaction of localized plasmons and results in the coherent coupling of plasmon mode with A and B excitonic states. The decoherence time is observed to be a few ps although the exciton dynamics are modified within a ps. This coupling modifies the exciton binding energy as the spacing between the excitonic bands is significantly reduced during the coherent interaction. There is a relatively higher density of excitons generated in $MoS_2$ with 2.33 eV excitation that gives rise to the exciton-exciton interaction in $MoS_2$. A simple density matrix model provides a qualitative description of the observed experimental results. For more quantitative description, one has to take into account the many-body effect and exact band structures of interacting states.

The excitation tuned to the C exciton absorption band at 3.06 eV only excites the carriers in $MoS_2$ to C excitonic state. The electrostatic screening in the presence of Ag nanoislands results in the enhanced dissociation of C excitons into free carriers, followed by the relaxation of these free carriers to A and B excitonic absorption states. There is a stronger photoinduced absorption in $MoS_2$ with the excitation at 3.06 eV due to resonant excitation of carriers to the C excitonic state. A red shift of the excitonic bands in the hybrid Ag-$MoS_2$ system is observed under both excitations due to band gap renormalization in the presence of excess carriers from the Ag-NPs.

The enhanced optical absorption in the hybrid structure under optical pumping at plasmon energy can manipulate the carrier relaxation to the excitonic absorption states at significantly lower input powers. The strong coupling among the excitons due to localized plasmons in the semiconductors can be utilized for the development of integrated photonic switches, modulators, and coherent optical systems in the nanoscale dimensions.

**EXPERIMENTAL PROCEDURES**

Large area monolayer $MoS_2$ of high quality were synthesized by chemical vapor deposition (CVD). Molybdenum Dioxide ($MoO_2$) (99% Sigma Aldrich) is sulfurized at a temperature of 750

$^0$C and atmospheric pressure using Sulfur powder (99.5% Alfa Aesar).[49] Ag layer with a thickness of 10 nm was grown over monolayer MoS$_2$ by the physical vapor deposition (PVD) method using a thermal evaporator at a base pressure of 10$^{-6}$ Torr. The Ag film thus formed was annealed in a CVD chamber under Argon atmosphere at a pressure of 5 torrs and a flow rate of 80 standard cubic centimeters per minute. This thermal process causes the nucleation of Ag thin film resulting in the formation of Ag nanoislands. When the Ag layer is annealed for 30 minutes at a temperature of 200 $^0$C, Ag nanoislands are formed over the MoS$_2$ monolayer, as shown in figure 1 (a) and (b). The metallization process that we used in this study is used by other groups as well and Ag does not react with monolayer MoS$_2$ under the growth conditions we used.[3,50]

The exciton dynamics were studied using pump-probe spectroscopy by measuring the transient absorption spectrum and decay kinetics. A 100 fs Ti: Sapphire oscillator seeded optical parametric amplifier laser was used for the optical excitation and a white-light probe was utilized for studying the broadband absorption of the excitonic states. The difference in absorption, which is defined as $\Delta A(\lambda) = -log \frac{I(\lambda)pumped}{I(\lambda)unpumped}$, is measured at different delay times by varying the distance traveled by the pump and probe and the corresponding transient spectrum at different delay times are analyzed to study the exciton dynamics in monolayer MoS$_2$ and the influence of Ag-NPs on the dynamics as early as fraction of ps using an optical delay line with two different pump energies – 2.33 eV and 3.06 eV. The optical excitation due to the pump at 2.33 eV is resonant to the localized plasmons generated within the Ag-NPs located on the surface of the MoS$_2$ layer and couples with the phonon modes. The optical pump at 3.06 eV is resonant to the C excitonic state in the MoS$_2$ and is detuned from the plasmon energy of Ag-NPs.

**SUPPORTING INFORMATION**

Transient absorption spectrum of MoS$_2$ and Ag-MoS$_2$ with 3.06 eV excitation, the spectral peak fitting analysis of the transient absorption spectra, the theoretical details of the density matrix model

*Acknowledgement*:

AN would like to acknowledge the financial support from AMMPI seed research project. SK would like to acknowledge financial support from NSF ECCS 1351424.

**For Table of Contents Use Only**

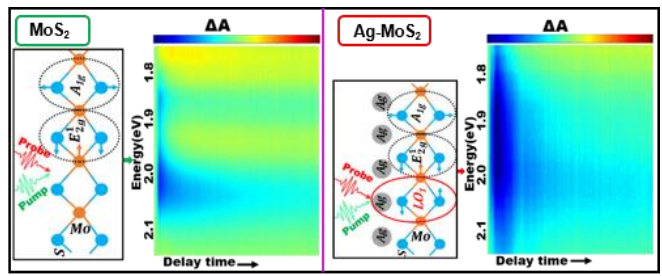